\documentclass[aps, %
twocolumn, %
preprintnumbers, %
nofootinbib, %
floatfix, %
superscriptaddress, %
]{revtex4} %
\IfFileExists{srcltx.sty}{\usepackage[active]{srcltx}}

\usepackage{amssymb}%
\usepackage{amsmath}%
\usepackage{graphicx}%
\usepackage{bm,paralist,xspace}

\IfFileExists{hypernat.sty}{\usepackage{hypernat}}

\usepackage[colorlinks=true,bookmarks=false]{hyperref}

\let\jnlstyle=\rm \def\jref#1{{\jnlstyle#1}} 
 \def\apj{\jref{ApJ}} \def\apjl{\jref{ApJ}}
   
\def\aap{\jref{A\&A}}   
  
 \def\mnras{\jref{MNRAS}}
 
 \def\prd{\jref{Phys.~Rev.~D}}
 
 \def\pasj{\jref{PASJ}}

 \def\nat{\jref{Nature}}

\newcommand{\fov}{{\mathrm{fov}}} 
\newcommand{\eff}{{\mathrm{eff}}} 
\newcommand{\vir}{{\mathrm{vir}}} 
\newcommand{\xmm}{{\emph{XMM-Newton}}\xspace} 
\newcommand{\dm}{{\textsc{dm}}} 
\newcommand{\ev}{\:\mathrm{eV}} 
\newcommand{\kev}{\:\mathrm{keV}} 
\begin{document}

\title{Search for the light dark matter with an X-ray spectrometer}

\author{A.~Boyarsky}%
\affiliation{CERN, PH/TH, CH-1211 Geneve 23, Switzerland} %
\affiliation{\'Ecole Polytechnique F\'ed\'erale de Lausanne, Institute of
  Theoretical Physics, FSB/ITP/LPPC, BSP 720, CH-1015, Lausanne, Switzerland}
\affiliation{\emph{On leave from} Bogolyubov Institute of Theoretical Physics,
  Kyiv, Ukraine}
\author{J.-W. den Herder}%
\affiliation{SRON, The Netherlands Institute for Space Research, Sorbonnelaan 2, 3854 CA Utrecht, the Netherlands} %
\author{A.~Neronov}%
\affiliation{INTEGRAL Science Data Center, Chemin d'\'Ecogia 16, 1290 Versoix,
  Switzerland and Geneva Observatory, 51 ch. des Maillettes, CH-1290 Sauverny,
  Switzerland}
  
\author{O.~Ruchayskiy}%
\affiliation{Institut des Hautes \'Etudes Scientifiques, Bures-sur-Yvette,
  F-91440, France}
\preprint{CERN-PH-TH/2006-227}%
\begin{abstract}
  Sterile neutrinos with the mass in the keV range are interesting warm dark
  matter (WDM) candidates. The restrictions on their parameters (mass and
  mixing angle) obtained by current X-ray missions ({\it XMM-Newton} or {\it
    Chandra}) can only be improved by less than an order of magnitude in the
  near future. Therefore the new strategy of search is needed.  We compare the
  sensitivities of existing and planned X-ray missions for the detection of
  WDM particles with the mass $\sim 1-20$~keV.  We show that existing
  technology allows an improvement in sensitivity by a factor of 100.  Namely,
  two different designs can achieve such an improvement: [A] a spectrometer
  with the high spectral resolving power of 0.1\%, wide (steradian) field of
  view, with small effective area of about cm$^2$ (which can be achieved
  without focusing optics) or [B] the same type of spectrometer with a smaller
  (degree) field of view but with a much larger effective area of
  $10^3$~cm$^2$ (achieved with the help of focusing optics).  To illustrate
  the use of the ``type A'' design we present the bounds on parameters of the
  sterile neutrino obtained from analysis of the data taken by an X-ray
  microcalorimeter.  In spite of the very short exposure time (100 sec) the
  derived bound is comparable to the one found from long {\it XMM-Newton}
  observation.
\end{abstract}

\maketitle

\section{Sterile neutrino as WDM candidate.}
\label{sec:wdm}

During the last year a number of works appeared, devoted to search for the
decay signal of a DM candidate -- sterile neutrino -- in the X-ray spectra of
astrophysical objects~\cite{Boyarsky:05,Boyarsky:06b,Boyarsky:06c,Riemer:06,
  Watson:06,Riemer:06b,Boyarsky:06d,Boyarsky:06e,Abazajian:06b}.  Indeed, it
was noticed long ago~\cite{Dodelson:93} that a right-handed neutrino with its
mass in the keV range presents a viable \emph{warm dark matter} (WDM)
candidate.  Such a particle would possess a specific radiative decay channel
and therefore one can search for its decay line in the X-ray spectra of
astrophysical objects~\cite{Dolgov:00,Abazajian:01b}.

The recent spark of interest in the search for sterile neutrino DM has several
reasons.  First, for the direct search of a DM particle, a particle physics
model is needed. Most of the particle physics candidates (axion,
supersymmetric particles, etc.) would constitute cold dark matter (CDM). CDM
models have several difficulties which could be resolved by a warm DM with the
particle mass in the keV range. In particular, WDM can ease the problem of the
dark halo structures in comparison with the CDM
scenario~\cite{Bode:00,Goerdt:06,Strigari:06}.  Second, as the Standard Model
of particle physics (SM) does not contain a DM candidate, most of the
extensions of the SM ( like, for example, supersymmetry) require to assume the
existence of many new particles and/or validity of new fundamental principles.
Such extensions are not based on any available experimental data, but on
theoretical arguments only.  From this point of view, the extension of the SM
with several right-handed neutrinos (i) is, maybe, a minimal extension of the
SM one can imagine; (ii) is based on experimental data; (iii) provides
naturally a \emph{warm} DM candidate. 

Indeed, the existence of right-handed neutrino particles would
provide the most natural explanation of neutrino oscillations, which cannot be
explained within the Standard Model (see e.g.~\cite{Strumia:06} for a review
of neutrino oscillations).  Adding three right-handed (or \emph{sterile})
neutrinos, neutral with respect to all the Standard Model (SM) gauge
interactions, makes the fermion sector of the SM fully symmetric, as every
left-handed fermion obtains a right-handed counterpart.  It has been
demonstrated recently in Refs.~\cite{Asaka:05a,Asaka:05b} that the parameters
of these right-handed particles can be chosen in such a way that this model
resolves another problem of the SM -- explains the excess of baryons over
antibaryons in the Universe (the \emph{baryon asymmetry}). At the same time it
does not spoil the predictions of the Big Bang nucleosynthesis and accommodates the
data on neutrino oscillations. For this to be
true, the masses of the two of these sterile neutrinos should be chosen in the
range $\mathrm{300\;MeV}\lesssim M_{2,3}\lesssim \mathrm{20\;GeV}$, while the
mass of the third (lighter) sterile neutrino is arbitrary (as long as it is
below $M_{2,3}$). As the coupling of this lightest right-handed particle with
ordinary matter can be made arbitrary weak (for example, it can entirely
decouple of the other SM particles, while being produced in the early Universe
via the decay of the inflaton~\cite{Shaposhnikov:06}), this particle provides a
viable DM candidate. For this its mass should satisfy the universal
Tremaine-Gunn lower bound~\cite{Tremaine:79}: $M_\dm\gtrsim 300-500\ev$. In
particular, its mass can be in keV range.

For other interesting
applications of sterile neutrinos with the mass in the keV range in
astrophysics and cosmology
see~\cite{Kusenko:06a,Biermann:06,Stasielak:06,Kusenko:06b,Hidaka:06}
and references therein.

The details of the process of structure formation in the Universe depend on the
mass of DM particles. In principle, comparison of the results of numerical
modeling of structure formation  with Lyman-$\alpha$ forest
data allows  to obtain a lower bound on the DM particle
mass~\cite{Seljak:06,Viel:06}. However, it turns out that such a bound is
model-dependent~\cite{Asaka:06,Viel:06}. This means that experimentally
interesting energy range for the search of the DM decay line is anywhere above
the lower limit determined by the Tremaine-Gunn bound, 
$E\gtrsim 150-250$~eV. 

If the mass of the sterile neutrino is less than the electron rest mass, it
can decay into a photon and an active
neutrino~\cite{Pal:81}.\footnote{Although throughout this paper we are talking
  mostly about the sterile neutrino, all the results can be applied to
  \emph{any} DM particle, possessing the monoenergetic radiative decay
  channel, emitting photon of energy $E_\gamma$ and having decay width
  $\Gamma$. For earlier works, discussing cosmological and astrophysical
  effects of decaying DM
  see~\cite{DeRujula:80,Berezhiani:87,Doroshkevich:89,Berezhiani:90a,Berezhiani:90b}.
  The extensive review of the results can be also in the
  book~\cite{Khlopov:97}.
} %
As the mass of the active neutrino is much smaller than 1 keV, the photon is
essentially monoenergetic: $E_\gamma = \frac{M_s}2$. The radiative decay width
$\Gamma$ is traditionally parameterized in terms of mass $M_s$ and mixing
angle $\theta$ -- measure of the interaction of the sterile neutrino with its
active counterparts. The radiative decay width is expressed via $M_s$ and
$\theta$ as~\cite{Pal:81,Barger:95}:
\begin{align}
  \label{eq:1}
  \Gamma &= \frac{9\, \alpha\, G_F^2} {1024\pi^4}\sin^22\theta\,
  M_s^5 \notag\\
  &\simeq 1.38\times10^{-22}\sin^2(2\theta)
  \left[\frac{M_s}{\mathrm{keV}}\right]^5\;\mathrm{sec}^{-1}\:.
\end{align}
where $\alpha$ is the fine-structure constant and $G_F=1.166\times
10^{-5}\;\mathrm{GeV}^{-2}$ is the Fermi coupling constant. The photon flux
from the DM decay is given by
\begin{equation}
\label{eq:3}
  F_\dm =  \frac{E_\gamma}{M_s}\;\Gamma\!\!\!\int\limits_{\fov\;
    \mathrm{cone}}\frac{\rho_\dm(\vec r)}{4\pi|\vec D_L + \vec r|^2}d^3\vec
  r\;. 
\end{equation}
Here $\vec D_L$ is the distance between the observer and the center of the
observed object, $\rho_\dm(r)$ is the DM density and the integration is over
the DM distribution inside the (truncated) cone -- solid angle, spanned by the
field of view (FoV) of a telescope.  If the observed object is
far\footnote{Namely, if luminosity distance $D_L$ is much greater than the
  characteristic scale of the DM distribution $\rho_\dm(r)$.},
Eq.~(\ref{eq:3}) simplifies to:
\begin{equation}
\label{eq:4}
F_\dm = \frac{M_\dm^\fov \Gamma}{4\pi D_L^2}\frac{E_\gamma}{M_s}\;,
\end{equation}
where $M_\dm^\fov$ is the mass of DM within a telescope's field of view (FoV).

Clustering of the DM at small red shifts results in the enhancement of the DM
decay signal in the direction of large nearby mass concentrations, such as the Milky
Way halo, nearby galaxies and galaxy clusters. Characterizing the clustering
scale through the typical overdensity compared to the mean cosmological DM
density $\rho_{\rm DM}^0$
\begin{equation}
{\cal R}=\rho/\rho_{\rm
  DM}^0
\end{equation}
(${\cal R}\sim 10^6$ for a galaxy, ${\cal R}\sim 10^3$ for a galaxy cluster)
and through its size, $D\sim 10$~kpc$\sim 10^{-5}H_0^{-1}$ ($H_0$ is the
Hubble constant) for a galaxy, $D\sim 1$~Mpc$\sim 10^{-3}H_0^{-1}$ for a
galaxy cluster, one can find that the DM decay flux from a particular
overdensity is comparable to the background DM decay signal. Indeed, in
Eq.(\ref{eq:4}) $M_{\dm}^{\fov}\simeq {\cal
  R}\rho^0_{\dm}DD_\theta^2\Omega_{\fov}$ and therefore
\begin{equation}
  \label{eq:2}
  \frac{F_ {\cal R}}{F_{\rm XRB}}\sim {\cal R}DH_0\sim 1\;.
\end{equation}

However, the spectra of the DM signal from the background and from a galaxy or
galaxy cluster are different.  The flux from a nearby object would be detected
as a Doppler-broadened line of the width
\begin{equation}
  \label{eq:6}
  \Delta E_\mathrm{line}=\frac{v_{\vir}}{c} E_{\gamma}
\end{equation} 
where $v_{\vir}$ is the virial velocity of the DM particles ($v_{\vir}\sim
10^{-2}c$ for a galaxy cluster, $v_{vir}\sim 10^{-3}c$ for a galaxy).  At the
same time, the DM decay contribution into XRB is produced by the decays at red
shifts $z\sim 0\div 1$ and, as a result the DM decay line is broadened to
$\Delta E\sim M_s/2$. Thus, in spite of the fact that the compact DM sources
at $z\simeq 0$ give just moderate enhancement of the DM decay flux, the
enhancement of the signal in the narrow energy band centered on the line
energy $E=M_s/2$ can be large for the \emph{instruments with high spectral
  resolution}.

\section{Sensitivity of X-ray telescopes for DM detection.}
\label{sec:sensitivity}

As it is discussed in the previous Section several types of astrophysical
objects are expected to produce comparable strength DM decay fluxes.
Galaxy clusters and nearby dwarf galaxies are extended sources of the size of
about $1^\circ$. The Milky Way halo is expected to produce a diffuse signal
detectable from all directions.

Most of the currently operating X-ray telescopes are not optimized for the
study of diffuse emission and/or very extended sources. For example, the field
of view of {\it Chandra} and {\it XMM-Newton} are, respectively,
256~$\mathrm{arcmin}^2$ and 700~$\mathrm{arcmin}^2$, much smaller than the
angular size of the typical DM dominated sources.  

To improve the existing bounds on DM
parameters~\cite{Boyarsky:05,Boyarsky:06b,Boyarsky:06c,Riemer:06,Watson:06,
  Riemer:06b,Boyarsky:06d,Boyarsky:06e,Abazajian:06b} using these instruments,
one can look for the sources with optimized ``signal-to-noise'' ratio (as e.g.
dwarf galaxies~\cite{Boyarsky:06c}) and study them with prolonged
observations.  However, as we have discussed in the previous section, the
expected DM signal does not vary too much from object to object. Even if an
object is undetected in X-rays, the bounds are defined by statistical error
which behaves as $\sqrt{t_\mathrm{exp}}$.  Therefore, improving the results by
an order of magnitude requires an exposure time two orders of magnitude
longer.  Thus, further improvement of the results will be very slow and would
require a lot of observational time of existing satellites.

\begin{figure}
  \includegraphics[width=\linewidth]{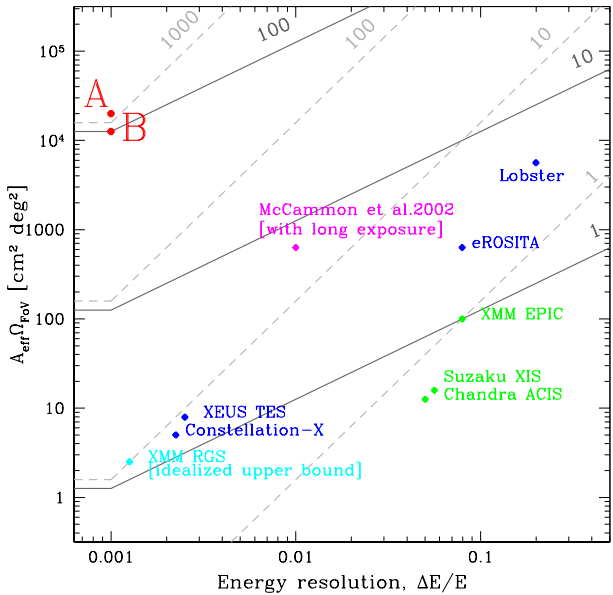}
\caption{Comparison of sensitivities of existing and  proposed/planned X-ray
  missions for the detection of the DM decay line from the Milky Way DM halo. 
  The plot shows the
  characteristics of different telescopes in the two parameter space relevant
  for the diffuse DM line detection: ``Energy resolution'' vs. ``Grasp''.  The sensitivity of {\it XMM-Newton} is taken as a reference.
  Solid lines limit from below the regions of parameter space in which
  the improvement of sensitivity by factors of 1, 100 and 100 can be
  achieved (each line is marked by the corresponding numerical factor). Dashed 
  lines show the improvement of conservative ``total flux'' bounds on neutrino
  mixing angle (see text for explanations) in the case of non-detection of the
  DM decay line. }
\label{fig:sensitivity} 
\end{figure}

%
A qualitative improvement of the bounds on the DM parameters (or final
detection of the DM decay signal) can be achieved only with a qualitatively
new instrumentation. To find what experimental set up can be considered as
``optimized'' for the DM search, it is useful to compare the sensitivities of
existing and planned X-ray missions for the detection of the DM decay signal.

To this end one needs to distinguish two possible situations -- \emph{strong
  line} and \emph{weak line} regimes.  If the line is \emph{weak} (i.e. there
is no line detectable against the continuum at more than several $\sigma$) and
the background signal can be fitted well by a convincing physical model (or
just a featureless power law model), detectabilty of the line is roughly
defined by the statistical error of the background flux in the corresponding
energy bin.  As the background count rate in a narrow energy band, centered at
the line energy, is proportional to the spectral resolution of the instrument
$\Delta E$, the significance of line detection scales as the inverse of the  
square root of spectral resolution
\begin{equation}
  \label{eq:8}
  {\rm LINE\ SIGNIFICANCE} \sim \frac{1}{\sqrt{1+\Delta E/ 
\mathrm{EW}}}\;.
\end{equation}
For a given total flux in the line, the sensitivity for the line  
detection improves as
\begin{equation} 
\label{lineflux}
{\rm SENSITIVITY} \sim\frac{1}{\sqrt{\Delta E}}
\end{equation} 
until the spectral resolution reaches the equivalent width (EW) of the line.
In this case one reaches the   ``strong line'' regime in which the detection
significance does not   depend on the statistical error of background anymore.
The relation~(\ref{lineflux}) holds until the spectral resolution becomes so
good that it either reaches the intrinsic width of the line~(\ref{eq:6}) (that
is, $\Delta E\sim 10^{-2}E$ for galaxy clusters and $\Delta E\sim 10^{-3}E$
for galaxies) or the background count rate in the narrow energy bin becomes so
small that one looks for a line signal in a background-free regime.

If the line is \emph{strong} 
the sensitivity
is defined by the intensity of the line itself. From this point of view,
making $\Delta E$ smaller could not just increase the sensitivity of an
instrument, but change the situation from the weak line to the strong line
regime and, therefore, increase the sensitivity much more significantly. The
condition for the line flux to exceed that of the background scales as
$1/\Delta E$.

Additionally, the search of the DM decay line in the X-ray energy band is
complicated by the fact that this line can be easily confused with atomic
emission lines present in the emission spectra of astrophysical plasmas.
Uncertainties of the models of diffuse emission from the warm and hot plasma
in the Galaxy prevent a proper statistical analysis of the data. This
difficulty is the main reason why many works, deriving  the bounds on the DM
sterile neutrino parameters, consider a simpler approach which gives weaker,
but more robust constraints. Namely, one simply requires that the DM decay
line flux in a given narrow energy band does not exceed the total background
flux in the same energy bin
(c.f.~\cite{Riemer:06,Boyarsky:06c,Watson:06,Boyarsky:06d,Riemer:06b}).
Although such approach does not allow to detect a line, if it is present in
the data, it permits to derive a ``background model independent'' bounds.

As explained above, within the ``total flux restrictions'' approach, the
bounds on the neutrino parameters, derived from the data, improve as
\begin{equation}
{\rm TOTAL\ FLUX\ BOUND}\sim \frac{1}{\Delta E}
\end{equation}
rather than as $1/\sqrt{\Delta E}$, as in the statistical analysis approach.
In Fig.~\ref{fig:sensitivity} we show with dashed grey lines the improvements
of the upper bound on the DM sterile neutrino mixing angle which can be
achieved within the "total flux restrictions" approach.  One can see that e.g.
for an instrument with the spectral resolving power of about $10^3$ and
collecting power of $10^4$~cm$^2$deg$^2$ the improvement can by by 3 orders of
magnitude, compared to the bounds derived from {\it XMM-Newton} data.

The imaging instruments on existing satellites have spectral resolution of
about 10\%.  However, there are grating spectrometra on board of both
\emph{XMM-Newton} and \emph{Chandra} telescopes.  The resolving power of
Reflection Grating Spectrometer (RGS) on board of \xmm is 10--40 times better
than that of EPIC cameras. Even if it were possible to use the full field of
view of this instrument with its maximal spectral resolution, this would
provide an order of magnitude improvement of sensitivity, as compared to the
EPIC camera of \xmm.  Such a spectral resolution is possible, however, only
for point sources.  For spatially extended objects the resolution degrades
proportionally to the angular size of the source (see
e.g.~\cite{Peterson:02,dePlaa:06}).  Taking into account that for the DM line
search one deals with extended sources (with the angular size larger than that
of the field of view of the RGS), one finds that the spectral resolution of
grating   instrument in this case is even worse than that of   EPIC camera.
%
To recover a better spectral resolution for an extended object, one must use
modeling, using a surface brightness profile of a given object.  For the DM
search, however, it should be noticed that the surface brightness profile for
the DM is much flatter than the surface brightness for the intracluster or
intergalactic media, as it is proportional not to the square of the density
profile, but to the density profile itself (as we are looking for the
one-particle decay process). This makes the effective size of the source
larger and should be taken into account in the corresponding data processing.
As a result, the existing gratings of XMM and Chandra can hardly improve the
sensitivity for the DM search significantly.

Microcalorimeters provide an alternative to the gratings in achieving high
spectral resolution.  The existing technology (c.f. e.g.~EURECA --
EURopean-JapanEse Calorimeter Array~\cite{McCammon:02,eureca:06,Kelley:06})
allows to construct an X-ray detector with 0.1\% spectral resolution at
several keV energies.

An obvious requirement for any instrument aimed at high-resolution
spectroscopic study is the maximal possible effective collecting area,
$A_{\eff}$, needed to increase the photon statistics in the narrow energy
bands corresponding to the spectral lines.  The statistics of the background
signal also grows with $A_{\eff}$ which leads to the fact that the sensitivity
scales only proportionally to the square root of the area
\begin{equation}
{\rm SENSITIVITY}\sim \sqrt{A_{\eff}}
\end{equation}
The need to maximize the effective area has pushed the design studies of the 
planned X-ray telescopes to consider the multi-spacecraft configuration in 
which a mirror with a large collecting area (5~m$^2$ in the case of {\it
XEUS}) flies separately from the detector.

However, if one is interested in the search of DM line from the Milky Way
halo, a simpler design allows to achieve sensitivity higher than the one that
would be reached with {\it XEUS}.  The point is that one can increase the photon
statistics by increasing the field of view of the telescope (even the ``wide
field'' camera on board of \emph{XEUS} is supposed to have a tiny field of
view of just 6 arcmin). For the case of the MW halo the line and background
photon statistics are just proportional to the size of the FoV,
$\Omega_{\fov}$, which means that the sensitivity for the DM line detection
scales as
\begin{equation}
{\rm SENSITIVITY}\sim \sqrt{\Omega_{\fov}}
\end{equation}

We compare the sensitivities of existing and future missions for the detection
of the DM decay line signal from the Milky Way halo on the
FIG.~\ref{fig:sensitivity}, taking sensitivity of \xmm\ EPIC camera as the
reference. The resolving power is plotted along the X-axis of
FIG.~\ref{fig:sensitivity}.  Since the overall photon statistics is
proportional to the product $A_{\eff}\Omega_{\fov}$ it is convenient to range
the X-ray missions according to their ``grasp'', $A_{\eff}\Omega_{\fov}$ .
This parameter is plotted along the Y axis in FIG.~\ref{fig:sensitivity}. The
diagonal solid lines, marked by the corresponding numbers, show the relative
improvement in sensitivity for different missions, as compared to
\emph{XMM-Newton}. The turnover of the lines at the resolving power $\Delta
E/E\sim 10^{-3}$ is related to the fact that further improvement of the
spectral resolution will not lead to the improvement of sensitivity, because
one hits the natural width of the line, determined by the Doppler broadening
due to the random motions of the DM particles in the gravitational potential
of the Galaxy. In Fig.~\ref{fig:sensitivity} the ``XMM RGS'' point (dark green
dot) marks the idealized upper bound for the grating if one could have used
its full field of view with maximal spectral resolution. As discussed above,
to use gratings in case of extended objects, one needs to model surface
brightness profile of both DM and gas components. As one sees from
FIG.~\ref{fig:sensitivity}, even this, hard to achieve, setup gives at most
moderate improvement compared to {\it XMM-Newton}/EPIC.  Shown in magenta
color is the spectrometer used by~\citet{McCammon:02} (see
Section~\ref{sec:mccammon} below). This spectrometer flew on a sounding rocket
and therefore had very short exposure time ($\sim 100$~sec).  Plotted on
FIG.~\ref{fig:sensitivity} is the sensitivity of the corresponding device if
it were placed on a satellite (in which case it would have much longer
exposure).  This explains, why the results of Section~\ref{sec:mccammon} below
do not provide an order of magnitude improvement, compared to the \xmm data
(c.f.  FIG.~\ref{fig:exclusion}).

\begin{figure}
  \includegraphics[width=\linewidth]{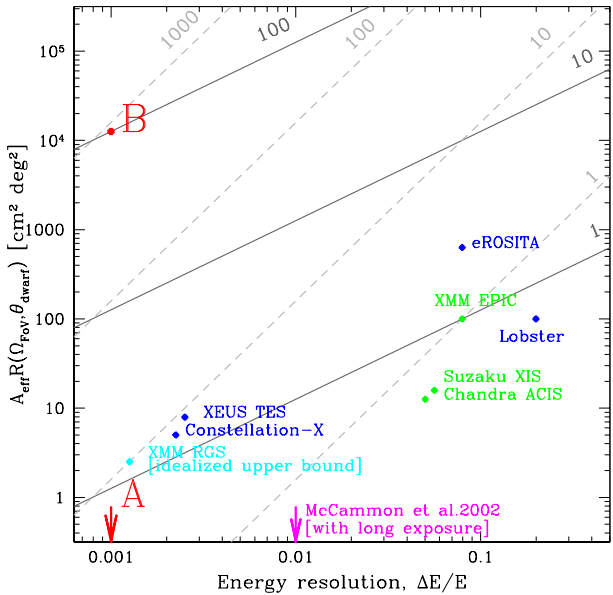}
\caption{Same as in Fig. \ref{fig:sensitivity}, but for the sensitivity for the
detection of DM decay line from a nearby dwarf galaxy of the angular size of 2
degrees.}
\label{fig:sensitivity_dwarf} 
\end{figure}

Fig. \ref{fig:sensitivity_dwarf} shows a comparison of sensitivities of
existing and planned missions for the detection of DM decay line from the
nearby dwarf galaxies (we take as a reference a galaxy of the angular size of
$\sim 2^\circ$). The main difference with the case of DM decay line from the
Milky Way halo is that the objects have finite angular extent. This means that
extending the field of view of an instrument improves the sensitivity only
when the size of the field of view is less or equal the angular size of the
dwarf galaxy, $\theta_{\rm dwarf}$. The $Y$ axis of Fig.
\ref{fig:sensitivity_dwarf} shows, therefore the product of effective area of
an instrument on a function
\begin{equation}
R(\Omega_{\rm FoV},\theta_{\rm dwarf})=\left\{
\begin{array}{ll}
\Omega_{\rm FoV}, & \mbox{ if } \Omega_{\rm FoV}\le
\pi\theta_{\rm dwarf}^2\\
\pi\theta_{\rm dwarf}^2, & \mbox{ if } \Omega_{\rm FoV}>
\pi\theta_{\rm dwarf}^2
\end{array}
\right.
\end{equation}
Another main difference between Fig. \ref{fig:sensitivity_dwarf} and 
Fig. \ref{fig:sensitivity} is that the lines representing the improvement of
sensitivity by a factor of 10 and 100 do not have a turnover at $\Delta
E/E\simeq 10^{-3}$, because the velocity dispersion of particles moving in the
halo of dwarf galaxy is an order of magnitude less than the velocity dispersion
of DM particles in the Milky Way halo. Thus, the Doppler broadening of the line
is not observable until the spectral resolution becomes $\Delta E/E\sim
10^{-4}$.

\begin{figure}
  \includegraphics[width=\linewidth]{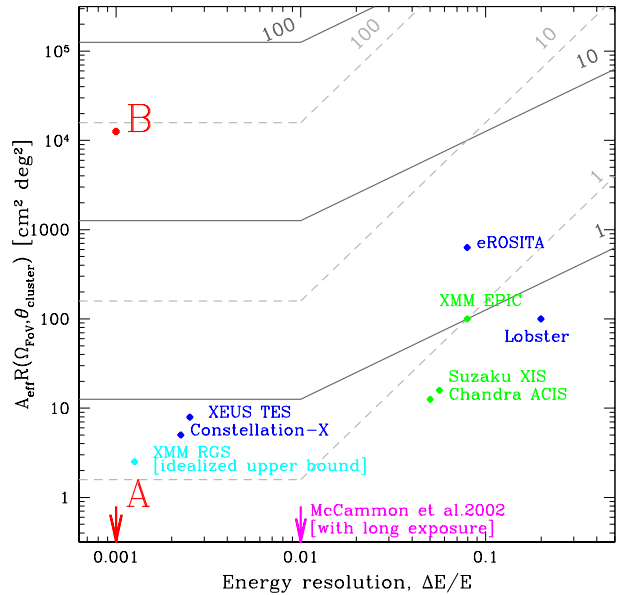}
\caption{Same as in Fig. \ref{fig:sensitivity}, but for the sensitivity for the
detection of DM decay line from a galaxy cluster of the angular size of 2
degrees.}
\label{fig:sensitivity_cluster} 
\end{figure}

Finally, Fig. \ref{fig:sensitivity_cluster} shows comparison of sensitivities of
different missions for the detection of the DM decay line from galaxy clusters.
One can see that because of the larger velocity dispersion of DM particles in
the galaxy clusters the improvement of the spectral resolution does not lead
anymore to the improvement of sensitivity already when the resolution $\Delta
E/E\sim 1\%$ is reached. Even disregarding the fact that the detection of DM decay line from
the galaxy clusters is complicated by the fact that clusters are strong sources
of continuum and atomic line X-ray emission, one can see from Fig.
\ref{fig:sensitivity} that no significant improvement of the bounds imposed by
{\it XMM-Newton} observations \cite{Boyarsky:06b} can be achieved with and of
existing and proposed experiments.

\section{Optimal experimental set-up.}
\label{sec:optimal-setup}

From Figs. \ref{fig:sensitivity}, \ref{fig:sensitivity_dwarf} one can clearly
see  what qualitative improvements of experimental set up are needed to 
improve  sensitivity for the DM decay line detection. If one aims at the
detection of the DM decay line from the Milky Way halo, one can choose two
different ways for such improvements, which we describe as ``type A'' and ``type B''
designs below. 

First of all, it is clear that an improvement of spectral resolution down to the
theoretical ``ultimate'' resolution of $\Delta E/E\sim 0.1\%$ enables to improve
the sensitivity by a factor of 10 compared to the sensitivity of {\it
XMM-Newton}. A microcalorimeter-type detector can achieve this spectral
resolution. Further improvements of the sensitivity can be achieved by
increasing the ``grasp'' of the instrument.

In the ``type A'' design, one maximizes the ``grasp'' by increasing
the size of the field of view up to (several) steradian. 
The wide field of view of the
instrument makes the use of focusing optics impossible and the only way to
maximize the effective area is to increase as much as possible the geometrical
area of the detector. In Fig. \ref{fig:sensitivity} we tentatively assume
that  an area of  3~cm$^2$ (10 times larger than the detector of
\citet{McCammon:02}) is reachable for a microcalorimeter-type detector
(although this can be challenging task with an existing technology). 

In the ``type B'' design one maximizes the ``grasp'' by increasing the effective
collection area of the detector with the help of the imaging optics.
In this case moderately extending the field of view to
 $1.5^\circ\times 1.5^\circ$, 
and increasing the effective collection area to
$1000~\mathrm{cm}^2$ will enable to  improve the {\it XMM-Newton} 
total flux restrictions by three orders of
magnitude and  to increase the sensitivity for the line detection by more than
an order of magnitude, provided that a microcalorimeter type detector is
installed in the focal plane.

One should note that both types of designs have their limitations. For
example, the ``type A'' set-up will be difficult to use at the energies below
1~keV because the DM decay signal will be contaminated by the strong
foreground emission from the local hot bubble, which contains a huge amount of
atomic emission lines. In this case the lack of proper imaging capabilities
would make the disentanglement of the DM decay line signal from the forest of
atomic lines extremely difficult. On the other side, in the ``type B'' set-up
the use of the focusing optics essentially limits the size of the field of
view, especially at the energies above 2~keV (higher energies require longer
focal length and hence larger detector area). Thus at the energies above
several keV ``type B'' design quickly becomes technically not feasible with
the current day technology.

Apart from the DM search the wide field of view both types of design can serve
for spectrometer for several engineering and scientific tasks. A focusing
X-ray telescope with a $1.5^\circ$ field of view and excellent spectral
resolution (``type B'' design) can serve for a variety of astrophysical
problems. The flight of the wide field of view spectrometer (``type A''
design) can be considered as a test of the X-ray spectrometers of the next
generation ``big'' X-ray mission, such as \emph{XEUS} or \emph{Constellation
  X}.  A slightly modified version of the ``bare spectrometer'' design would
also make soft X-ray measurements of the prompt emission from gamma-ray bursts
(GRBs).  This is a particularly interesting task in the view of the claimed
detection of transient line and/or edge-like features in the X-ray spectra of
prompt emission of several GRBs~\cite{Amati:00,Frontera:01}.

In fact, none of the proposed X-ray mission is optimized to verify this claim, 
because the detection of X-ray lines and edges in the spectra of prompt
emission of GRBs requires a wide field of view X-ray detector.  Up to now only
the Wide Field Camera (WFC) on board of {\it BeppoSAX} mission satisfied this
requirement (the field of view of about $40^\circ$ and moderate spectral
resolution of $\sim 20$\%). The WFC has detected the spectral features
(probably associated with iron) in 5 GRBs~\cite{Amati:00,Frontera:01}.
Sensitivity  of the wide field of view spectrometer for the line detection 
will be comparable to the sensitivity of the WFC, because the decrease of the
effective area (by a factor of 20) will be compensated by the gain in the
spectral resolution by a factor of 200.

Although the ``type A'' and ``type B'' designs have comparable sensitivity for
the detection of the DM decay line from the Milky Way halo, the sensitivity of
the wide field of view design for the detection of DM decay signal from the
nearby dwarf galaxies is extremely low, as one can see from Fig.
\ref{fig:sensitivity_dwarf}. However, this does not make the ``type B'' design
preferable compared to the ``type A'' one. Indeed, as it is mentioned above,
the ``type B'' instruments can operate only in a narrow energy range
$0.1-(several)$~keV, while the energy region interesting for the sterile
neutrino DM search extends definitely above 1-2~keV energies. At higher
energies the wide field of view spectrometer becomes the only available
configuration which will provide an increase of sensitivity by a factor of 100
compared to the sensitivities of existing instruments.

\section{Analysis of the X-ray spectrometer data.}
\label{sec:mccammon}

A ``prototype'' of a cryogenic X-ray spectrometer has already been
successfully tested in flight~\cite{McCammon:02}. The detector was composed of
36 micro-calorimeters of the surface area of 1~mm$^2$ each and had a wide,
1~sr field of view.  In this Section we will show that, although the flight
time of this calorimeter was short, about 100 seconds, good characteristics of
the detector (spectral resolution of 10~eV) allow this instrument to compete
with {\it XMM-Newton} in restricting parameters of sterile neutrino DM.

We have analyzed the data obtained by the X-ray spectrometer\footnote{We thank
  Prof.~D.~McCammon for providing us the data for the analysis} in order to
derive the upper bounds on the neutrino mixing angle as a function of the
mass.\footnote{When this work was finished, the paper~\cite{Abazajian:06b}
  appeared, which compared estimates for the neutrino line flux in the
  Dodelson-Widrow model at $E=1\kev$ with the flux, measured
  by~\cite{McCammon:02} at the same energy.}
At the energies below keV the signal collected by the detector is dominated by
the diffuse X-ray background of which some fraction is produced by distant
active galactic nuclei~\cite{Mushotzky:00,Worsley:04,Treister:05} and can be
modeled with an absorbed power-law, while the rest is due to the thermal
emission from the local hot bubble (the temperature $T\simeq 0.1$~keV). The
thermal emission is dominated by the atomic emission lines. The X-ray
background spectrum measured by the X-ray spectrometer and the identifications
of the detected atomic lines can be found in Ref.~\cite{McCammon:02}.

The presence of bright emission lines in the background spectrum complicates
the search of the DM decay line. The main problem is that the DM can ``hide''
behind an atomic line if the energies of two lines are close enough. The
problem of ``hiding behind the line'' can, in principle, be relaxed if one
properly models the background thermal emission: the ratios of the intensities
of the multiple emission lines from the same element are fixed if the
temperature of the gas is known.  The DM decay line would then reveal itself
if an ``anomalous'' ratio between line intensities is found at a particular
energy. In the statistical analysis of the data, addition of the DM decay line
at the right energy on top of the thermal emission model would then improve
the quality of the model fit to the data by removing the ``anomaly'' in the
line intensity ratio.

The statistical analysis outlined above has sense  at large enough signal to
noise ratios. However, because of the short exposure time of the X-ray
spectrometer under consideration, the brightest lines were detected at
$3-4\sigma$ level and statistical analysis gives the results comparable to the
results obtained with a more simple analysis method which does not depend on
the assumptions about the background emission model.

The background model independent bound on the neutrino parameters can be
derived from the fact that the flux in a narrow line at a given energy can not
exceed the total detected flux (possible line plus background) in a narrow
energy band of the width equal to the energy resolution of the instrument and
centered at the line energy. We have ``scanned'' the whole energy interval of
interest, $0.2$~keV$<E<1$~keV, calculating the maximal allowed flux at
$3\sigma$ level in a narrow energy window of the width 10~eV, equal to the
energy resolution of the spectrometer. With such a method one does not have a
possibility to really detect a DM decay line, one can only obtain a robust
restriction on the DM parameters.

\begin{figure}[t]
  \includegraphics[width=\linewidth]{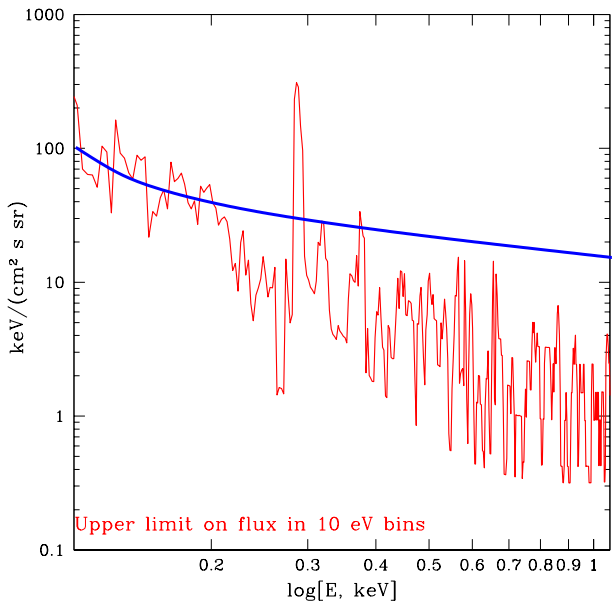}
\caption{ Red thin line: an upper limit on the flux in the dark matter decay
line as a function of energy. Blue thick line: average level (line features
smoothed) of soft X-ray spectrum in the 0.1-1~keV energy range.} 
\label{fig:spectrum} 
\end{figure}

The limit on the total flux in 10~eV energy bins as a function of energy is
shown in FIG.~\ref{fig:spectrum} with the red solid line. One can see that the
upper limit on the total flux is weaker at the energies where real atomic
lines are present. The X-ray photon flux is, in general, affected by the
photo-electric absorption in the interstellar medium of the Galaxy. In order
to correct the data for the effect of photoelectric absorption, one has to
know the hydrogen column density in the direction of the source. In the case
of interest, the source (the DM particles decaying in the Galactic halo) is
distributed all over the Galaxy and a correct calculation of the effect of
absorption should involve an integration of absorption of the DM signal along
the line of sight.  The result of such a calculation would depend on the
assumptions about the model of spatial distribution of the hydrogen in the
Galaxy. In order to impose a conservative upper limit on the possible DM line
flux one can use a simplified procedure which slightly over-estimates the
effect of photo-electric absorption, namely, one can assume that the signal is
absorbed on the total Galactic hydrogen column density $N_H(l,b)$ in a given
direction. Using the known distribution of the hydrogen column density
\cite{Dickey:90}, we find that the mean hydrogen column density throughout the
spectrometer field of view in the pointing direction of the instrument
($l=+90^\circ, b=+60^\circ$ in Galactic coordinates) is $N_H\simeq 1.5\times
10^{20}$~cm$^{-2}$. FIG. \ref{fig:spectrum} shows the data corrected for the
absorption on this hydrogen column density.

\begin{figure}[t]
  \includegraphics[width=\linewidth]{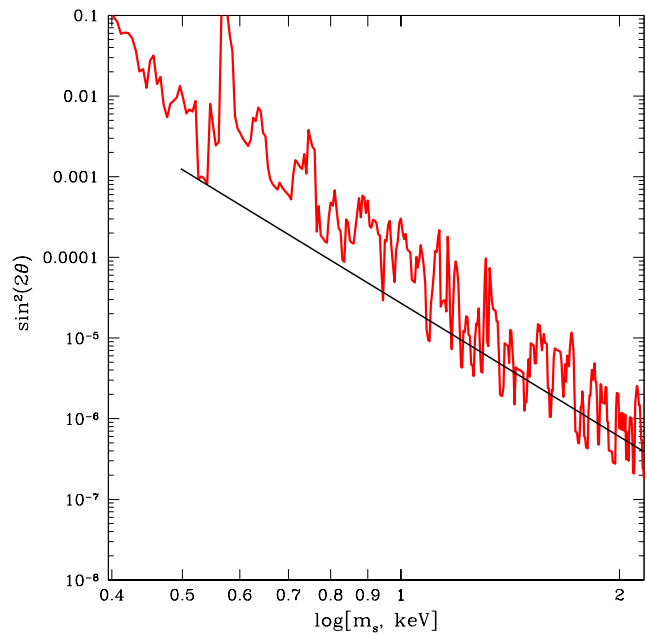}
\caption{Exclusion plot in the $(m_s,\sin^22\theta)$ parameter space obtained
  from the analysis of the X-ray spectrometer data of~\citet{McCammon:02}
  (solid thick red line). For comparison, the upper limit on the mixing angle
  obtained from the {\it XMM-Newton} observation of the Large Magellanic Cloud
  obtained in \cite{Boyarsky:06c} is shown by the solid thin black line.  }
\label{fig:exclusion} 
\end{figure}

In order to derive the upper bound on the neutrino mixing angle as a function
of the neutrino mass from the above upper limit on the DM decay line flux, one
has to calculate the expected DM decay line flux within the 1~sr field of view
of the detector. The field of view of the detector was pointed in the
direction $l=+90^\circ, b=+60^\circ$ in Galactic coordinates.  The expected DM
decay line flux (Eq.~(\ref{eq:3})) in the case of Galactic DM halo was
computed in~\cite{Boyarsky:06c}. It varies from its minimum value,
$F_{\dm,\mathrm{min}}$ from the direction of Galactic anti-center to the
maximum, $F_{\dm,\rm max}\simeq 6F_{\dm,\rm min}$ from the direction of the
Galactic center. For a conservative upper limit we can assume that the flux
from the direction of the North Galactic Pole is about the minimal flux
$F_{\dm,\rm min}$. Relating via Eq.~(\ref{eq:1}) the DM decay width $\Gamma$
to parameters of sterile neutrino DM $M_s$ and $\sin^2(2\theta)$ we obtain:
\begin{equation}
  \label{eq:7}
  F_\dm = 3.84\times 10^4\sin^2(2\theta)
  \left[\frac{M_s}{\mathrm{keV}}\right]^5\;\mathrm{\frac{keV}{cm^2\cdot
      sec\cdot sr}}
\end{equation}
The emitted photon has the energy $E_\gamma = M_s/2$.
The bound on the neutrino mixing angle as a function of neutrino mass derived
from the upper limit on the flux in 10~eV energy bins
(FIG.~\ref{fig:spectrum}) is shown in Figure \ref{fig:exclusion}. For
comparison we show in the same figure the bound derived from the statistical
analysis of the {\it XMM-Newton} data \cite{Boyarsky:06c}.  A remarkable fact
is that in spite of the very short exposure time (100~s compared to 18~ksec
exposure in {\it XMM-Newton} LMC data) the bounds derived from the data of the
two instruments are comparable. (Moreover, the bound from the X-ray
spectrometer is even gets better than that of {\it XMM-Newton} for some DM
masses $M_s\ge 1$~keV (i.e with the line energy above 0.5~keV)). 
This demonstrates the advantages of the ``type  
A'' experimental set-up, outlined above: the large field of view
makes the ``grasp'' of an instrument even with a small effective area  
(like the spectrometer under consideration) comparable to the ``grasp''
of the focusing instrument with a small field of view (like {\it XMM- 
Newton}). Combined with a good spectral resolving power, such design  
turns out to be better suited for the DM line search than {\it XMM- 
Newton}.


\section{Conclusion}
\label{sec:conclusion}

In this paper we analyzed existing and planned X-ray missions with respect to
their ability to detect signal of radiatively decaying DM. Our analysis shows
that two types of experimental setups enable to improve the sensitivity for
the detection of the DM decay line by a factor of 100 compared to the existing
instruments, such as {\it XMM-Newton} and {\it Chandra}. Namely, a high
spectral resolution detector with resolving power $\Delta E/E\sim 10^3$
incorporated either in a wide (steradian scale) field of view telescope with
limited imaging capabilities (``type A'' design) or into a focusing telescope
with much smaller (degree scale) field of view, but with a much larger
collection area (``type B design'') can achieve such an improvement. Both type
A and type B designs are optimized for the search of the DM decay signal from
the DM halo of the Milky Way galaxy. The ``type B'' design will also be able
to search the DM decay signal from the nearby dwarf galaxies and from the
galaxy clusters.  The two designs are complimentary to each other because the
``type B'' design is optimized for the search of DM signal in the energy range
from 0.2~keV up to 1-2~keV, while the ``type A'' design is more suitable for
the search of DM signal at the energies above 1~keV (and can be extended to
10--15~keV).

\subsection*{Acknowledgements}
\label{sec:acknowledgement}

We would like to thank D.~McCammon for sharing the data with us and
T.~Maccarone, M.~Markevitch, M.~Shaposhnikov, I.~Tkachev and J.~M.~in~'t~Zand
for stimulating scientific discussions and for providing very useful comments
on the manuscript. The work of A.B. was (partially) supported by the EU 6th
Framework Marie Curie Research and Training network "UniverseNet" (MRTN-
CT-2006-035863).  The work of O.R. was supported in part by the European
Research Training Network contract 005104 ``ForcesUniverse'' and by a
\emph{Marie Curie International Fellowship} within the $6^\mathrm{th}$
European Community Framework Programme.


\end{document}